# Ultra-low lattice thermal conductivity in tungsten-based scheelite ceramics

Hicham Ait Laasri [a,*], Eliane Bsaibess [a,b], Fabian Delorme [a], Guillaume F. Nataf [a], Fabien Giovannelli [a,*]

[a] GREMAN, Université de Tours – CNRS – INSA Centre Val de Loire - UMR7347, IUT de Blois, 15 rue de la chocolaterie, CS32903- 41029 Blois cedex, France

* Corresponding authors: fabien.giovannelli@univ-tours.fr, hicham.aitlaasri@univ-tours.fr

[b] Present address: Sciences and Engineering Department, Sorbonne University Abu Dhabi, Al Reem Island, PO Box 38044, Abu Dhabi, United Arab Emirates

**Abstract**

$BaWO_4$, $Ce_{2/3}\square_{1/3}WO_4$ and $La_{2/3}\square_{1/3}WO_4$ polycrystalline ceramics were synthesized by conventional solid-state reaction route. The effect of cation-deficiency on the crystallographic structure, microstructure and thermal properties of these scheelite-type compounds were investigated. X-ray diffraction was used to identify the single-phase scheelite structure of the studied ceramics. Scanning Electron Microscopy technique has revealed a homogenous and dense microstructure with a few micro-cracks. The thermal conductivity of $BaWO_4$ scheelite decreases from $1.3\pm0.2$ to $1.0\pm0.1$ W m$^{-1}$ K$^{-1}$ in the range 373 K – 673 K. The cation-deficient scheelites $Ce_{2/3}\square_{1/3}WO_4$ and $La_{2/3}\square_{1/3}WO_4$ ceramics display an ultra-low thermal conductivity of $0.3\pm0.04$ W m$^{-1}$ K$^{-1}$ and $0.2\pm0.03$ W m$^{-1}$ K$^{-1}$ at 673 K, respectively. These materials exhibit among the lowest known values of thermal conductivity in crystalline oxides, in this temperature range. Therefore, they appear as very attractive for thermal barrier coating and thermoelectric applications.

# 1. Introduction

The discovery of materials with low thermal conductivity offers great opportunities for many applications [1] such as thermal barrier coatings (TBCs) [2-5] and solid-state thermoelectric converters [6-7]. With the extensive demand for more efficient and powerful aero-engines, the development of advanced thermal barrier coatings has attracted impressive attention to find suitable thin oxide-ceramics (1µm – 5µm), particularly useful for gas turbine blades to thermally insulate air-cooled metallic components from hot gases in the engine, enhancing their combustion efficiency, performance, and longevity [2].

Nowadays, 7-8 wt. % yttria-stabilized zirconia (YSZ) is the main successful ceramic oxide used as thermal barrier coating due to its impact resistance, low thermal conductivity (∼1.5-3 $W\ m^{-1}\ K^{-1}$) [4], high thermal expansion, high melting point and excellent chemical stability [5]. Nevertheless, its structural phase transition at temperatures above 1473 K increases the risk of crack propagation [8-9]. To improve the performances of TBCs, several scientific and technical approaches have been adopted to discover material with low thermal conductivity. Previous studies have reported that some inorganic materials, such as rare earth zirconates [10], pyrochlore oxides [11] or $La_2Mo_2O_9$-based compounds [5, 12] exhibit a lower thermal conductivity than YSZ [4]. Recently, low thermal conductivity has been reported in high entropy oxides [13-15] or guided by probe structure and machine learning [16]. Besides, due to the development of global economy, the demand for energy is considerably increasing for industrial production and human life.

Thermoelectric materials, which can directly convert waste heat into electricity, are the subject of numerous studies [17]. High thermoelectric performance can be achieved by increasing the power factor and reducing the thermal conductivity of materials [5-6, 18-19]. In this research field, many different routes have been explored to decrease the thermal conductivity of promising materials such as alloying, nano-structuring, composites, microcracking approaches or improving electrical properties of intrinsic low thermal conductivity materials [5-6, 18-26]. Increasing crystal complexity is one of these approaches to obtain ultra-low lattice thermal conductivity in inorganic materials [12, 18, 27-28]. Previous studies focused on complex crystal structures and materials that have a large unit cell volume and a large molecular weight, due to the large population of diffuson-like modes [12, 18, 27-28]. Another promising approach to reduce the thermal conductivity is to introduce cation vacancies [29-34]. For example, in perovskites ($ABO_3$), the introduction of La-site vacancies in $SrTiO_3$ decreases the thermal conductivity from 10 W m$^{-1}$ K$^{-1}$ to ~2 W m$^{-1}$ K$^{-1}$ and results in a glass-like behavior [29, 33].

Searching for novel thermally insulating materials, first principles calculations and high-throughput calculations have predicted an ultra-low thermal conductivity of $ABO_4$ scheelite-type structure, composed into the stacking of weak units [$AO_8$] and strong units [$BO_4$] [35-36]. In addition, the strong bonds favour a high temperature stability, while the weak bonds lead to large thermal expansion coefficients [37] and good damage tolerance [35-36, 38]. In particular, the low Pugh's ratio of scheelites indicate that they are quasi-ductiles [35-36]. They are thus good candidates as TBCs.

Recently, the predictions of ultra-low thermal conductivities have been confirmed by Bsaibess *et al*. [39], who have reported a value around 1 W m$^{-1}$ K$^{-1}$ from 400 to 1000 K in $BaMoO_4$. They have also demonstrated that the presence of vacancies on the A-site, in the $La_{2/3}\square_{1/3}MoO_4$ molybdate scheelite, leads to even lower thermal conductivity values of about 0.6 W m$^{-1}$ K$^{-1}$ over the entire temperature range [400 – 1000 K] [39], in agreement with theoretical findings [36]. Based on the same calculation approach, $BaWO_4$ should exhibit a lower thermal conductivity (0.8 W m$^{-1}$ K$^{-1}$) than $BaMoO_4$ (1 W m$^{-1}$ K$^{-1}$) [36]. Therefore, in this study, we investigate the sintering and thermal properties of $BaWO_4$ ceramics and A-site deficient scheelites $Ce_{2/3}\square_{1/3}WO_4$ and $La_{2/3}\square_{1/3}WO_4$.

## 2. Experimental procedure

BaWO$_4$, Ce$_{2/3}$□$_{1/3}$WO$_4$ and La$_{2/3}$□$_{1/3}$WO$_4$ ceramics were synthesized by conventional solid-state reaction method using a stoichiometric amount of high purity precursors of BaCO$_3$ (99.9%), WO$_3$ (99.9%), CeO$_2$ (99,9%) from Sigma Aldrich and La$_2$O$_3$ (99.9%) from ChemPur. La$_2$O$_3$ was pre-heated at 1000 °C for 10 hours to remove water from this very hygroscopic precursor. Appropriate amounts of these precursors were then weighed and ground in planetary tungsten carbide ball mill at 300 rpm for 1 hour (Retsch PM 100). The obtained powders were subsequently calcined in air using alumina crucibles in a muffle furnace at 1173 K during 4 h for BaWO$_4$ and Ce$_{2/3}$□$_{1/3}$WO$_4$, and 1273 K during 10 h for La$_{2/3}$□$_{1/3}$WO$_4$. The calcined powders were then ground with few drops of polyvinyl alcohol (PVA) solution (2 wt. % in water) and pressed into pellets of 10 mm in diameter and about 2 mm in thickness under a pressure of 125 MPa. The obtained pellets were heated thereafter on zirconia beads in an alumina crucible at a heating rate of 3 K min$^{-1}$, up to 773 K and then a heating rate of 1 K min$^{-1}$, up to 873 K to drive off the PVA, and sintered for 4 hours at 1573 K, 1273 K and 1323 K for BaWO$_4$, Ce$_{2/3}$□$_{1/3}$WO$_4$ and La$_{2/3}$□$_{1/3}$WO$_4$, respectively. Relative densities were calculated from the pellets by both geometrical measurements (mass and dimensions) and the Archimedes method using a MS204TS/00 analytical balance (Mettler Toledo) and the theoretical density values obtained by structural refinement.

X-ray diffraction (XRD) analysis was performed at room temperature on the ceramic pellets using a Bruker D8 Advance diffractometer system, with λ=1.5418 Å powered at 40 kV × 40 mA, and a scanning step of 0.02° (2θ) per second. The obtained XRD patterns were refined using JANA software in order to confirm the single-phase scheelite-structure.

Microstructures were inspected by Scanning Electron Microscopy (TESCAN MIRA 3 SEM), employing secondary electrons (SE) with an acceleration voltage of 5 kV on the fracture of ceramics covered with an ultra-thin layer of gold to avoid charging effects. The average grain size of ceramics is estimated by the linear intercept method using the image J software. Elementary compositions were analysed by Energy Dispersive X-ray (EDX) spectroscopy, using the scanning electron microscopy (Tescan Mira 3 SEM). The homogeneity of ceramics was confirmed by back-scattering electron (BSE) analyses.

Specific heat capacity ($C_p$) of all samples was measured by Differential Scanning Calorimetry (DSC, Netzsch STA 449 F3 Jupiter) technique. The crushed ceramics (about 50 mg) were placed in a platinum crucible and heated continuously up to 900 K under nitrogen atmosphere with a heating rate of 20 K min$^{-1}$.

Thermal diffusivity measurements of ceramics were performed by the Laser Flash method (Netzsch LFA 457 instrument), under vacuum ($10^{-2}$ mbar). All obtained samples were previously coated with a thin-layer of graphite to improve absorption of the laser light and avoid emissivity errors.

## 3. Results and discussion

The XRD patterns at room temperature of $BaWO_4$, $Ce_{2/3}\square_{1/3}WO_4$ and $La_{2/3}\square_{1/3}WO_4$ ceramics are presented in Fig. 1a. All the obtained peaks can be attributed to the pure scheelite-type phase of tungsten oxides. The diffraction peaks of $BaWO_4$ reveal a tetragonal symmetry with the space group I41/a, indexed according to ICDD files (PDF cards 01-072-0746). The $Ce_{2/3}\square_{1/3}WO_4$ and $La_{2/3}\square_{1/3}WO_4$ diffractograms crystallize in a monoclinic symmetry with the space group C2/c [40-41]. The collected data of both compounds are in good agreement with ICDD files (PDF cards 01-085-0143 and 01-082-2068, respectively). Fig. 1b, 1c and 1d show the experimental and calculated XRD patterns of $BaWO_4$, $Ce_{2/3}\square_{1/3}WO_4$ and $La_{2/3}\square_{1/3}WO_4$, respectively, as well as their difference. The lattice parameters, volumes of the unit cell, theoretical and experimental densities are presented in table 1.

The Archimedes relative densities of the sintered ceramics are ≈ 97% for $BaWO_4$ and ≈ 93% for $Ce_{2/3}\square_{1/3}WO_4$ and $La_{2/3}\square_{1/3}WO_4$. Relative densities obtained from the mass and dimensions of the sintered ceramics are ≈93%, and ≈ 90% for $Ce_{2/3}\square_{1/3}WO_4$ and $La_{2/3}\square_{1/3}WO_4$. The difference between both methods arises from the open porosity of the ceramics. Microstructure study and grain size distribution of fractured surfaces of $BaWO_4$, $Ce_{2/3}\square_{1/3}WO_4$ and $La_{2/3}\square_{1/3}WO_4$ ceramic pellets imaged by Scanning Electron Microscopy are shown in Fig. 2. The micrographs show a homogeneous microstructure and a well-developed grain morphology for all specimens. This is consistent with the measured relative density of about 97% for $BaWO_4$ and 93% for $Ce_{2/3}\square_{1/3}WO_4$ and $La_{2/3}\square_{1/3}WO_4$. Distinct grain boundaries are observed in all samples. From histograms, the grain size of the scheelite ceramics is distributed in the range of ~ 5–50 µm. SEM images reveal average grain sizes of ~24 µm for $BaWO_4$ ceramic and ~20 µm for $Ce_{2/3}\square_{1/3}WO_4$ and $La_{2/3}\square_{1/3}WO_4$ ceramics. Moreover, for the three ceramics, both pores and cracks are observed. The EDX analysis gives cation composition ratios of 0.98 (Ba/W) in $BaWO_4$ and 0.68 (Ce/W and La/W) in $Ce_{2/3}\square_{1/3}WO_4$ and $La_{2/3}\square_{1/3}WO_4$. EDX measurements confirm thus the expected compositions for the three studied ceramics, within a 2% accuracy. Back-scattering electron images show that there is no chemical contrast for the three ceramics, confirming the good homogeneity and distribution of elements.

Fig. 3 shows the measurements of specific heat capacity for $BaWO_4$, $Ce_{2/3}\square_{1/3}WO_4$ and $La_{2/3}\square_{1/3}WO_4$ crushed ceramics as a function of temperature, from 330 to 900 K. The specific heat of $BaWO_4$ increases with increasing temperature from 0.34 J $g^{-1}$ $K^{-1}$ at 330 K to 0.46 J $g^{-1}$ $K^{-1}$ at 900 K. Ran *et al.* reported similar specific heat capacity for a $BaWO_4$ single crystal: from 0.32±0.01 J $g^{-1}$ $K^{-1}$ to 0.36±0.01 J $g^{-1}$ $K^{-1}$ between 336 K and 573 K [42]. Specific heat capacity of $Ce_{2/3}\square_{1/3}WO_4$ and $La_{2/3}\square_{1/3}WO_4$ have been measured for the first time. $C_p$ of $Ce_{2/3}\square_{1/3}WO_4$ varies from 0.34±0.01 J $g^{-1}$ $K^{-1}$ at 330 K to 0.46±0.02 J $g^{-1}$ $K^{-1}$ at 900 K. $C_p$ of $La_{2/3}\square_{1/3}WO_4$ varies from 0.32±0.01 J $g^{-1}$ $K^{-1}$ at 330 K to 0.42±0.02 J $g^{-1}$ $K^{-1}$ at 900 K. They are thus similar to the values found for $BaWO_4$, within the ±4% precision of heat capacity measurements.

The thermal diffusivity ($\lambda$) measurements of BaWO$_4$, Ce$_{2/3}\square_{1/3}$WO$_4$ and La$_{2/3}\square_{1/3}$WO$_4$ ceramics are shown in Fig. 4a. The thermal diffusivity of the three samples decreases when the temperature is increased from 0.588, 0.166 and 0.109 mm$^2$ s$^{-1}$ at 373 K to 0.369, 0.106 and 0.081 mm$^2$ s$^{-1}$ at 773 K for BaWO$_4$, Ce$_{2/3}\square_{1/3}$WO$_4$ and La$_{2/3}\square_{1/3}$WO$_4$, respectively. This decrease is due to the Umklapp scattering [43]. The thermal diffusivity is slightly increasing at 873 K by 18% for Ce$_{2/3}\square_{1/3}$WO$_4$ and 15% for La$_{2/3}\square_{1/3}$WO$_4$, while it is constant for BaWO$_4$. This is attributed to an increase of the radiative contribution at high temperatures, as reported in previous studies using the laser flash method [5]. Overall, it can still be seen that the A-deficient scheelites Ce$_{2/3}\square_{1/3}$WO$_4$ and La$_{2/3}\square_{1/3}$WO$_4$ exhibit very low thermal diffusivity values in the temperature range.

The thermal conductivities ($\kappa$) of BaWO$_4$, Ce$_{2/3}\square_{1/3}$WO$_4$ and La$_{2/3}\square_{1/3}$WO$_4$ ceramics are shown in Fig. 4b. Thermal conductivity is determined from measurements of thermal diffusivity ($\lambda$), specific heat ($C_p$) and density ($\rho$) through the following formula:

$$\kappa = \lambda \cdot \rho \cdot C_p \tag{1}$$

The measurement of thermal conductivity was carried out three times on each studied ceramics to confirm repeatability of the results. The standard deviation has been calculated, which is about 0.02 W m$^{-1}$ K$^{-1}$.

The thermal conductivity of all samples diminishes with increasing temperature between 373 K and 673 K. It then slightly increases, most likely because of an enhanced radiative contribution that is known to play a role in laser flash measurements [5]. BaWO$_4$ thermal conductivity decreases from 1.3±0.2 W m$^{-1}$ K$^{-1}$ to 1.0±0.1 W m$^{-1}$ K$^{-1}$, which corresponds to the theoretical prediction of Liu *et al.* [36]. However, these values are lower than those reported by Ran *et al.* on a BaWO$_4$ single crystal, which are 2.1 and 1.6 W m$^{-1}$ K$^{-1}$ at respectively 373 and 550 K [42]. This difference could be linked to the porosity and to the presence of microcracks and grain boundaries in the sample. To consider the impact of porosity, thermal conductivity values were normalized to represent values for 100% dense materials ($\kappa_{dense}$) using Maxwell's relation [5] as follow:

$$\kappa_{100\% \, dense} = \kappa_{measured} \times \frac{1}{1-1.5\Phi} \tag{2}$$

Where $\Phi$ is the porosity.

In Fig. 4b, these corrected values are plotted for BaWO$_4$ ceramic and still show a decrease of about 30% as compared to the single crystal. The thermal conductivity is 1.4±0.2 and 1.1±0.1 W m$^{-1}$ K$^{-1}$ at 373 and 550 K, respectively. This difference is attributed to the presence of grain boundaries that introduce an interface thermal resistance.

The thermal conductivity of Ce$_{2/3}\square_{1/3}$WO$_4$ decreases from 0.37±0.05 W m$^{-1}$ K$^{-1}$ to 0.30±0.04 W m$^{-1}$ K$^{-1}$ and that of La$_{2/3}\square_{1/3}$WO$_4$ is almost constant at about 0.2±0.03 W m$^{-1}$ K$^{-1}$ between 373 K and 673 K. These values are lower than in BaWO$_4$ and even lower than the calculated minimum thermal conductivity ($\kappa_{min}$) of BaWO$_4$ [35]. The minimum thermal conductivity can be expressed as a function of the number density of atoms and the average sound velocity, with three different models that relate to each other's as follow:

$$\kappa_{diff} = 0.72 \, \kappa_{Clarke} = 0.63 \, \kappa_{Cahill} \tag{3}$$

The main difference between these three models is that Clarke and Cahill assume propagating phonon-like vibrations, while $\kappa_{diff}$ is based on diffuson-like vibrations [44]. For BaWO$_4$, $\kappa_{Clarke}$ and $\kappa_{Cahill}$, have been already calculated based on first-principle calculations in ref. 35: $\kappa_{Clarke}$ = 0.61 W m$^{-1}$ K$^{-1}$, $\kappa_{Cahill}$ = 0.69 W m$^{-1}$ K$^{-1}$. These values lead to $\kappa_{diff}$ = 0.44 W m$^{-1}$ K$^{-1}$. The thermal conductivity experimentally measured for Ce$_{2/3}\square_{1/3}$WO$_4$ and La$_{2/3}\square_{1/3}$WO$_4$ scheelites are lower than the $\kappa_{min}$ calculated for BaWO$_4$ whatever the model used. However, a major difference with respect to BaWO$_4$ is the presence of a large amount of cation vacancy in these compounds that can strongly reduce the thermal conductivity, as shown previously for La$_{2/3}\square_{1/3}$MoO$_4$ deficient scheelite [39] and A-deficient ABO$_3$ perovskites [29, 32, 34].

Another approach consists thus in modelling the impact of vacancies. Indeed, the influence of vacancies on the thermal conductivity can be estimated by considering mass-difference scattering [45-47]. We consider here a fictive material $Ba_{2/3}\square_{1/3}WO_4$ and calculate the change in mass on the A-site induced by vacancies, such that:

$$\overline{\Delta M_A^2} = \frac{2}{3}(M_{Ba} - \overline{M_A})^2 + \frac{1}{3}(M_{Ba} + 2<\overline{M}>)^2 = 22306 \text{ g}^2 \text{ mol}^{-2} \qquad (4)$$

With $<\overline{M}>$ the average mass of the compound, $\overline{M_A}$ the stoichiometry weighted average of A-site average mass, and $M_{Ba}$ the molar mass of barium. This fictive material is close to $La_{2/3}\square_{1/3}WO_4$ and $Ce_{2/3}\square_{1/3}WO_4$ as molar masses of Ba, La and Ce are respectively 137.3, 138.9 and 140.1 g mol$^{-1}$. This leads to the average mass variance in the system $<\overline{\Delta M^2}>$ normalized by the squared average atomic mass $<\overline{M}>^2$ equal to 1.16. It is then possible to calculate the ratio of the defective solid's thermal conductivity ($Ba_{2/3}\square_{1/3}WO_4$) to that of a reference pure solid ($BaWO_4$) through Klemens model [48-49]. In our case, using values from first-principle calculations for the speed of sound and the volume per atom [35], and considering as a reference the thermal conductivity of $BaWO_4$ we measured at room temperature (1.3 W m$^{-1}$ K$^{-1}$), we find that the thermal conductivity of the compound with vacancies would be 0.50 W m$^{-1}$ K$^{-1}$, i.e. two times less than the thermal conductivity of the sample without vacancies. Following the same reasoning, we expect the thermal conductivity of $Ce_{2/3}\square_{1/3}WO_4$ and $La_{2/3}\square_{1/3}WO_4$ scheelites to be ultra-low. In this considered mass-difference scattering model, only kinetic perturbation due to mass difference is considered for the scattering parameter. However, vacancies lead also to a force constant difference and radius difference. In addition, in some cationic deficient oxides, it has been shown that the experimental thermal conductivity is lower than the one predicted by the mass-difference scattering only, and that the influence of force constant difference and radius difference must be taken into account as well [45]. In the case of $La_{1-x}CoO_{3-x}$, a value of ~0.6 W m$^{-1}$ K$^{-1}$ is measured for x = 0.1 [34]. The model with only mass-difference scattering gives a thermal conductivity of 1.6 W m$^{-1}$ K$^{-1}$, whereas the model with mass-difference and virial-theorem treatment for broken bonds, gives a thermal conductivity of 0.7 W m$^{-1}$ K$^{-1}$ which is very close to the measured one. If we assume that in scheelites, the rate of decrease is similar, it leads to a thermal conductivity of 0.2 W m$^{-1}$ K$^{-1}$ which is very closed to the measured thermal conductivity.

The thermal conductivity that we report for La$_{2/3}$□$_{1/3}$WO$_4$ is among the lowest lattice thermal conductivity values reported for a dense crystalline oxide ceramic. As already mentioned, in the case of BaWO$_4$, the influence of the microstructure accounts for about 30% of the decrease in thermal conductivity between the single crystal and the ceramic. If we assume a similar impact of the microstructure in La$_{2/3}$□$_{1/3}$WO$_4$, this leads to an ultra-low thermal conductivity around 0.3 W m$^{-1}$ K$^{-1}$ for a single crystal.

## 4. Conclusions

BaWO$_4$, Ce$_{2/3}$□$_{1/3}$WO$_4$ and La$_{2/3}$□$_{1/3}$WO$_4$ dense crystalline ceramics were prepared using the conventional solid-state reaction route. Dense ceramics have been obtained (97% for BaWO$_4$ and 93% for Ce$_{2/3}$□$_{1/3}$WO$_4$ and La$_{2/3}$□$_{1/3}$WO$_4$). XRD analysis revealed a single-phase scheelite-type structure for all ceramics. The crystal structure transformed from tetragonal for BaWO$_4$ to monoclinic for Ce$_{2/3}$□$_{1/3}$WO$_4$ and La$_{2/3}$□$_{1/3}$WO$_4$. The thermal measurements showed a low thermal conductivity for BaWO$_4$, below 1 W m$^{-1}$ K$^{-1}$ in the investigated temperature range [373 – 900 K]. In this work, ultra-low thermal conductivity values of 0.3±0.04 and 0.2±0.03 W m$^{-1}$ K$^{-1}$ at 673 K are reported for Ce$_{2/3}$□$_{1/3}$WO$_4$ and La$_{2/3}$□$_{1/3}$WO$_4$, respectively. These values are among the lowest thermal conductivities reported for crystalline oxides. The results of this study confirm that such A-deficiency is an efficient approach for lowering the thermal conductivity in scheelites. Combined with chemical doping to increase the electrical conductivity, it could lead to high-performance thermoelectric materials.

**Conflicts of Competing Interest**

The authors declare that they have no conflicts of interest.

**Acknowledgement**

The authors are grateful to Tatiana Chartier for technical support.

| Compound | BaWO$_4$ | Ce$_{2/3}$□$_{1/3}$WO$_4$ | La$_{2/3}$□$_{1/3}$WO$_4$ |
|---|---|---|---|
| a (Å$^3$) | 5.6288 | 7.8298 | 7.8757 |
| b (Å$^3$) | 5.6288 | 11.7408 | 11.8359 |
| c (Å$^3$) | 12.7281 | 11.6043 | 11.6539 |
| Cell volume (Å$^3$) | 403 | 1006 | 1025 |
| Theoretical density (g.cm$^{-3}$) | 6.3 | 6.75 | 6.61 |
| Geometrical relative density (%) | 93 | 90 | 90 |
| Archimedes relative density (%) | 97 | 93 | 93 |

**Table 1.** Refined lattice parameters, volume of the unit cell, theoretical and relative densities of sintered BaWO$_4$, Ce$_{2/3}$□$_{1/3}$WO$_4$ and La$_{2/3}$□$_{1/3}$WO$_4$ scheelite ceramics.

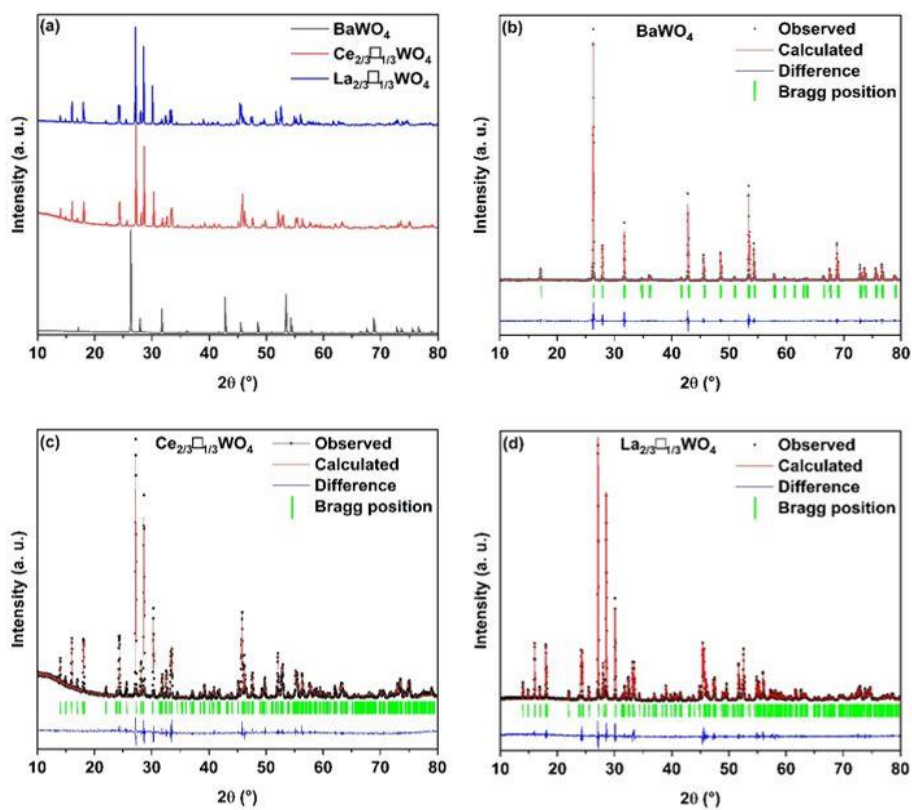

**Figure 1.** a) XRD patterns of BaWO$_4$, Ce$_{2/3}$□$_{1/3}$WO$_4$ and La$_{2/3}$□$_{1/3}$WO$_4$ ceramics at room temperature. Experimental and calculated XRD patterns of b) BaWO$_4$ c) Ce$_{2/3}$□$_{1/3}$WO$_4$ and d) La$_{2/3}$□$_{1/3}$WO$_4$.

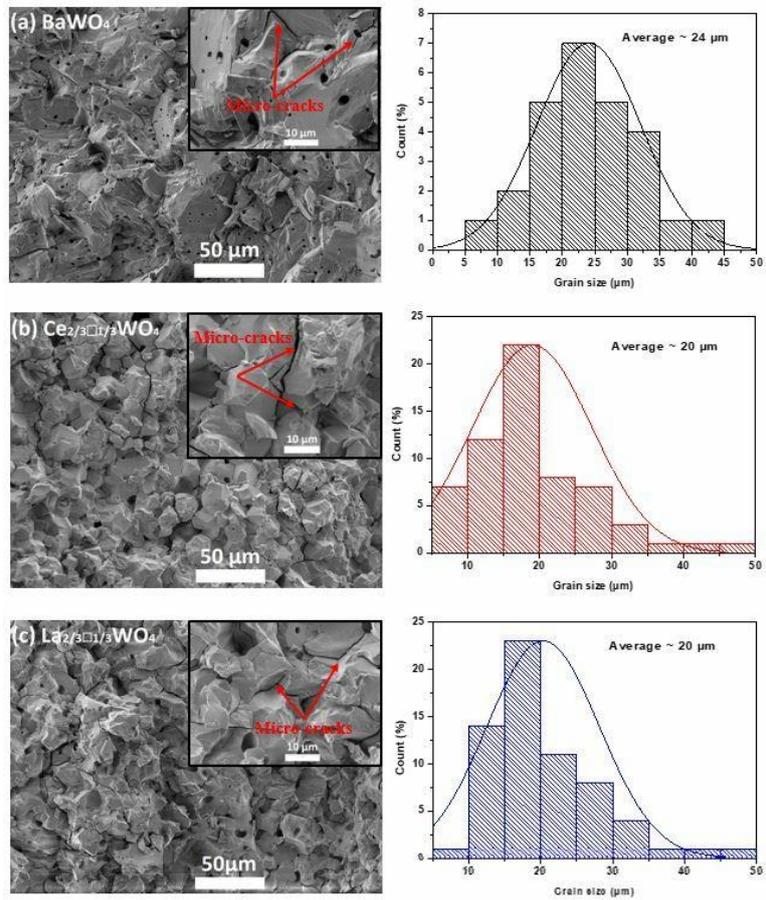

**Figure 2.** SEM micrographs and grain size distribution of a) BaWO$_4$, b) Ce$_{2/3}$□$_{1/3}$WO$_4$ and c) La$_{2/3}$□$_{1/3}$WO$_4$.

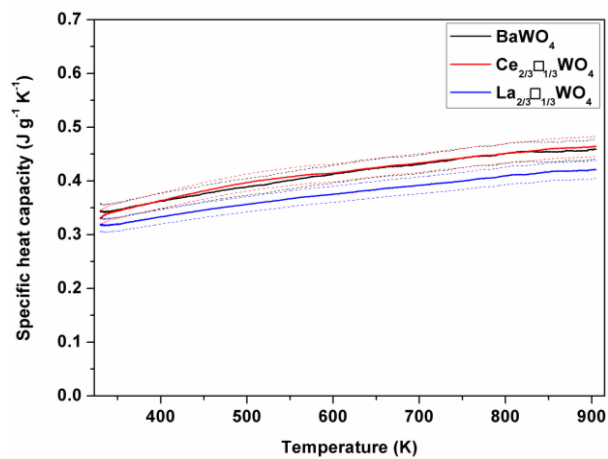

**Figure 3.** Temperature dependence of specific heat capacity $C_p$ of BaWO$_4$, Ce$_{2/3}$□$_{1/3}$WO$_4$ and La$_{2/3}$□$_{1/3}$WO$_4$. Dashed lines indicate the ±4% precision of heat capacity measurements.

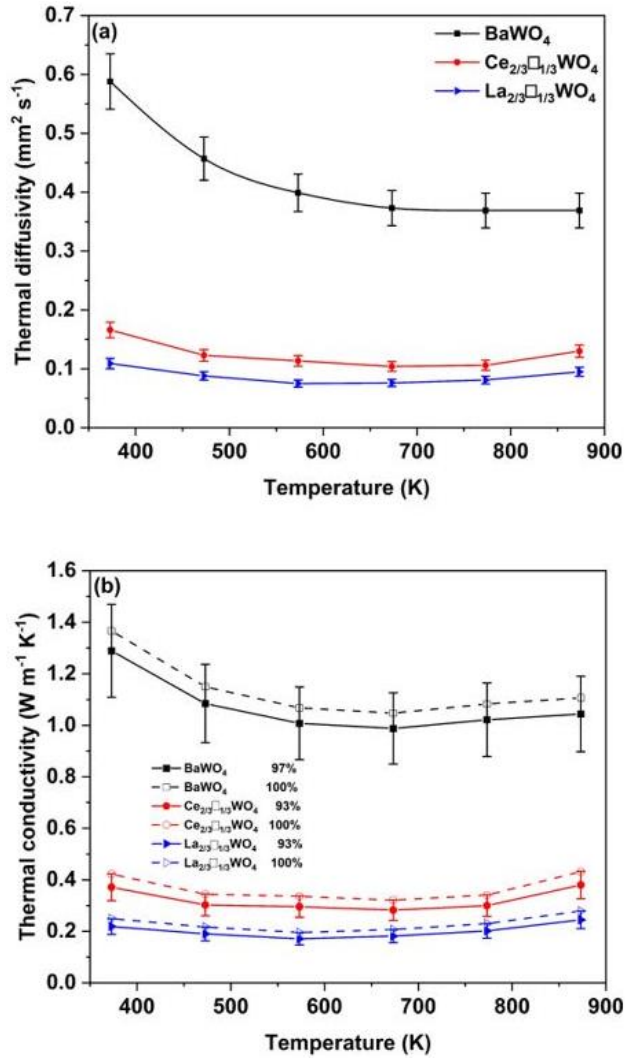

**Figure 4.** Temperature dependence of a) thermal diffusivity and b) thermal conductivity of BaWO$_4$, Ce$_{2/3}$□$_{1/3}$WO$_4$ and La$_{2/3}$□$_{1/3}$WO$_4$ ceramics. Values for 100% are corrected for the influence of the porosity. Error bars indicate the ±8% precision of thermal diffusivity measurements and ±14% precision on thermal conductivity data.